\documentclass[preprint2]{aastex}

\newcommand{\lsi}{LS~I~$+61~303$}
\newcommand{\ls}{LS~5039}

\slugcomment{Submitted to ApJ}
\shorttitle{Orbits of \lsi~and \ls}
\shortauthors{Aragona}

\begin{document}

\title{The Orbits of the $\gamma$-ray Binaries LS~I~+61~303 and LS~5039}

\author{Christina Aragona, M.\ Virginia McSwain}
\affil{Department of Physics, Lehigh University, 16 Memorial Drive East, Bethlehem, PA 18015}
\email{cha206@lehigh.edu, mcswain@lehigh.edu}

\author{Erika D.\ Grundstrom}
\affil{Physics and Astronomy Department, Vanderbuilt University, 6301 Stevens Center, Nashville, TN 37235}
\email{erika.grundstrom@vanderbilt.edu}
             
\author{Amber N.\ Marsh, Rachael M.\ Roettenbacher, Katelyn M.\ Hessler}
\affil{Department of Physics, Lehigh University, 16 Memorial Drive East, Bethlehem, PA 18015}
\email{anm506@lehigh.edu, rmr207@lehigh.edu, kmh410@lehigh.edu}

\author{Tabetha S.\ Boyajian}
\affil{Center for High Angular Resolution Astronomy and Department of Physics and Astronomy, Georgia State University, Atlanta, GA}
\email{boyajian@chara.gsu.edu}

\author{Paul S.\ Ray}
\affil{Space Science Division, Naval Research Laboratory, Washington DC, 20375}
\email{paul.ray@nrl.navy.mil}

\begin{abstract}
\lsi~and \ls~are two of only a handful of known high mass X-ray binaries (HMXBs) that exhibit very high energy emission in the MeV-TeV range, and these ``$\gamma$-ray binaries'' are of renewed interest due to the recent launch of the \textit{Fermi Gamma-ray Space Telescope}.  Here we present new radial velocities of both systems based on recent red and blue optical spectra.  Both systems have somewhat discrepant orbital solutions available in the literature, and our new measurements result in improved orbital elements and resolve the disagreements.  The improved geometry of each orbit will aid in studies of the high energy emission region near each source. 
\end{abstract}

\keywords{binaries: spectroscopic -- stars: individual(LS~5039, LS~I~$+61~303$) }

\section{Introduction}

\lsi~is a Be/X-ray binary with spectral type B0 Ve, a 26.5 day orbital period, and highly eccentric orbit \citep{grundstrom2007, casares2005a}.  High energy emission from \lsi~was first detected 30 years ago (as $\gamma$-ray source CG 135+1) by the Cosmic Ray Satellite B \citep{hermsen1977}.  Shortly thereafter, NASA's Einstein Observatory and ground based radio observations suggested that the system was no ordinary HMXB \citep{gregory1978, bignami1983}.  While the system has a relatively low X-ray luminosity for a HMXB, \lsi~is one of the 20 brightest $\gamma$-ray sources known.  The Be disk interacts with the compact companion, either a neutron star or black hole, producing orbital phase modulated emission across the electromagnetic spectrum:  TeV  \citep{albert2006, albert2008}, X-ray \citep{paredes1997, leahy2001}, optical H$\alpha$ \citep{grundstrom2007}, and radio flux \citep{gregory1978, taylor1992}. 

LS 5039 is a HMXB that consists of a massive, main sequence primary (ON6.5~V((f)) spectral type; \citealt{mcswain2004}) and a compact companion in a tightly bound, eccentric orbit \citep{casares2005b}.  Optical and UV spectra of the system reveal nitrogen enrichment and carbon depletion at the surface of the O star, a sign of prior mass transfer that inverted the binary mass ratio before the supernova occurred \citep{mcswain2004}.  The binary remained bound and was ejected from the Galactic plane within the past 1.1 Myr in an asymmetric supernova explosion that imparted a runaway velocity of about 150 km~s$^{-1}$ to the system \citep{ribo2002, mcswain2004}.  
As a HMXB, \ls~is quite inconspicuous with a relatively constant, low X-ray flux and hard spectrum; it does not experience dramatic X-ray flares or state changes that are seen in most other HMXBs.  However, it is an unusually bright radio source for a HMXB \citep{ribo1999}.  The system distinguished itself when it was first noted as a probable $\gamma$-ray source by \citet{paredes2000} since it is the only simultaneous X-ray and radio emitter inside the 95\% confidence contour of the EGRET source 3EG J1824$-$1514.  Its bright and hard $\gamma$-ray spectrum suggested that it might be observable by ground-based atmospheric Cherenkov telescope arrays, and indeed it was detected with energies $> 0.1$ TeV by the HESS array \citep{aharonian2005a}.  The identification of the $\gamma$-ray source with the optical star \ls~was strengthened by the $\gamma$-ray variability over the 4 day orbital period \citep{aharonian2006}.  The X-ray flux is also modulated with the orbital period \citep{bosch2005, bosch2007, takahashi2008}.  

\lsi~and \ls~are two of only five known HMXBs that exhibit very high energy emission in the MeV-TeV range (``$\gamma$-ray binaries''); the others are PSR B1259$-$63 \citep{cominsky1994, aharonian2005b}, Cygnus X-1 \citep{albert2007}, and the recently proposed candidate HESS J0632+057 \citep{hinton2009}.  The recent launch of the \textit{Fermi Gamma-ray Space Telescope} heralds a new era in the understanding of very high energy emission of these sources, and our group is currently involved in multiwavelength observing campaigns  for both systems.  In an effort to support these programs, we present here new optical spectroscopy and radial velocity measurements of \lsi~in \S 2 and \ls~in \S 3.  We discuss the results and the implications of the new orbits in \S 4.  


\section{Observations and Radial Velocity Measurements of \lsi}

We obtained 83 spectra of \lsi~at the KPNO coud\'e feed telescope over 35 consecutive nights during 2008 October and November.  We used grating B in second order with the OG550 order-sorting filter and the F3KB detector.  The spectra were wavelength calibrated using ThAr comparison lamp spectra taken throughout each night.  This instrumental configuration resulted in a wavelength range of 6400--7050 \AA~with a resolving power $R = \lambda/\Delta \lambda$ ranging from 11400--12500 across the chip.

We generally obtained 2--3 spectra of 30 minutes duration each night.  The images were zero corrected, flat fielded, and wavelength calibrated using standard procedures in IRAF\footnote{IRAF is distributed by the National Optical Astronomy Observatory, which is operated by the Association of Universities for Research in Astronomy, Inc., under cooperative agreement with the National Sciences Foundation}.  The spectra were interpolated onto a log wavelength scale using a common heliocentric wavelength grid, and they were rectified to a unit continuum using line-free regions.

\begin{figure}
\includegraphics[angle=90,scale=0.33]{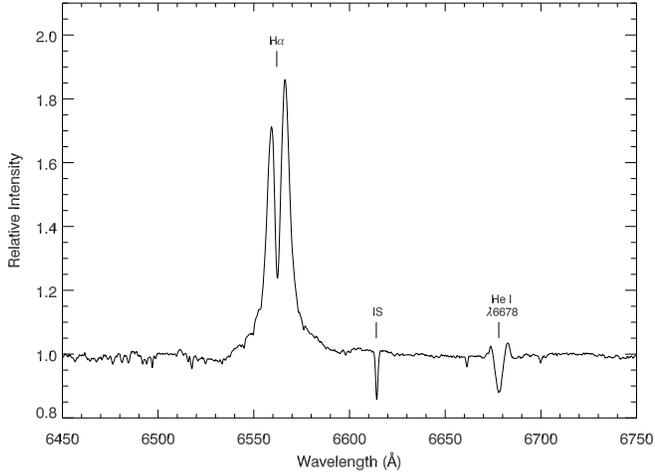} 
\\
\caption{The final mean spectrum of LS~I $+61~303$, formed by shifting each spectrum to its rest velocity to remove orbital variations.  The red spectrum reveals strong emission in the H$\alpha$ line and weaker emission in the wings of the \ion{He}{1} $\lambda6678$ line.  
\label{lsi_mean} }
\end{figure}

Our red spectra of \lsi~include only two prominent stellar features, H$\alpha$ and \ion{He}{1} $\lambda6678$, shown in Figure \ref{lsi_mean}.  The H$\alpha$ line is dominated by double-peaked emission from the circumstellar disk of the Be star.  While the absorption trough does follow the orbital motion of the star, we did not use this line to measure the spectroscopic orbit since its radial velocity, $V_r$, measurements may be strongly affected by changes in the disk structure.  Weak emission is also present in the blue and red wings of the \ion{He}{1} $\lambda6678$ line, but it is less affected by temporal changes in the disk.  Therefore we measured radial velocities of \lsi~using only this line.  

\placefigure{lsi_mean}

We formed a reference spectrum of LS~I~$+61~303$ using the mean of all available spectra (without shifting to their rest velocity to avoid any prior assumptions about the orbit), and we measured the $V_r$ of this mean spectrum using a parabolic fit of the \ion{He}{1} $\lambda6678$ line core.  Since we were concerned that the emission in the line wings (rather than the photospheric absorption component) might affect the relative velocities, we excised the emission on the blue and red wings of the reference spectrum to a unit continuum.  We then performed a cross correlation between each spectrum and this reference spectrum over the \ion{He}{1} $\lambda6678$ line region to obtain the relative $V_{r}$, which we added to the $V_r$ of the reference spectrum to obtain the absolute $V_r$.  We also measured the cross correlation velocity shift of the interstellar line at 6613 \AA, $\Delta V_{\rm ISM}$, using the same method to monitor the stability of our measurements.  After removing measurements with $|V_{\rm ISM}| > 5$ km~s$^{-1}$ and spectra in which the \ion{He}{1} $\lambda6678$ line profile was strongly affected by cosmic rays, we present a total of 75 $V_r$ measurements in Table \ref{vr}.  

\placetable{vr}

The orbital period of \lsi~is well known from radio flux observations \citep{gregory2002}, so we adopted the period $26.4960 \pm 0.0028$ d from that work.  Many recent papers also use an arbitrary orbital phase, $\phi (\rm TG)$, defined to be zero at HJD 2,443,366.775 \citep{taylor1982}, which we adopt as well.  Our $V_r$ agree well with \citet{casares2005a}, but measurements from \citet{grundstrom2007} are systematically lower by 8.8 km~s$^{-1}$ (based on the difference in the center of mass velocity, $\gamma$, using each data set separately).  This difference is likely due to a small difference in the rest wavelength of that line; we used 6678.15 \AA~as given by the NIST Atomic Spectra Database \citep{ralchenko2008}.  Therefore we offset their $V_r$ by this amount to determine the orbital elements of \lsi, and we excluded their points with discrepant $V_{\rm ISM}$ values as well.  Using this edited set of $V_r$ from \citet{casares2005a}, \citet{grundstrom2007}, and this work, we determined the orbital elements using the non-linear, least-squares fitting program of \citet{morbey1974}.  There are several points with large scatter and a systematic trend in the $V_r$ curve (discussed in \S4), so we assigned zero weight to points with $|O - C| > 2 \sigma$ and refit the orbit for the final orbital solution.  

The resulting time of periastron, $T_0$, eccentricity, $e$, angle of the line of nodes, $\omega$, velocity semiamplitude, $K_1$, center of mass velocity, $\gamma$, mass function, $f(m)$, projected separation, $a_1 \sin i$, and the standard deviation of the fit, $\sigma$, are listed in column 2 of Table \ref{orbit}.  The final $V_r$ curve is shown in Figure \ref{vrcurve_lsi}, and the orbital phases (relative to \citealt{taylor1982}) and observed minus corrected errors ($O-C$) are included in Table \ref{vr}.  With our new orbital fit, the orbital phase of periastron is $\phi (\rm TG) = 0.275$.

\placefigure{vrcurve_lsi}
\placetable{orbit}

\begin{figure}
\includegraphics[angle=90,scale=0.35]{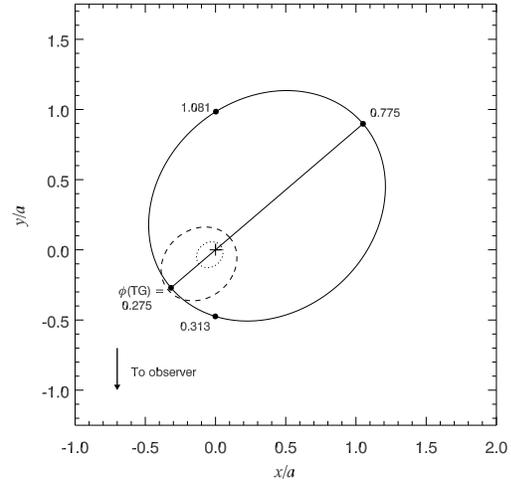} 
\\
\caption{Orbital geometry of \lsi, looking down on the orbital plane, showing the relative orbits ($r/a$) of the optical star ($12.5 M_\odot$; \citealt{casares2005a}) and its compact companion of unknown mass.  The relative orbit of the compact object is shown as a solid line, while the Be star's relative orbit depends greatly on the mass of the companion.  The dashed line indicates the Be star's orbit assuming a $4 M_\odot$ black hole, while the dotted line assumes a $1.4 M_\odot$ neutron star.   The relevant phases of periastron, apastron, and conjunctions are marked along the orbit of the compact companion.  The center of mass is indicated with a cross, and the thin solid line is the orbital major axis.
\label{lsi_orbit} }
\end{figure}

The geometry of the orbit is illustrated in Figure \ref{lsi_orbit}, assuming a Be star mass of $12.5 M_\odot$ \citep{casares2005a} with either a neutron star ($1.4 M_\odot$) or a black hole ($4 M_\odot$) companion.  
Except for $\gamma$, each orbital element agrees very well with the values found by \citet{grundstrom2007}.  Because our spectra of \lsi~were obtained over 35 consecutive nights, our $V_r$ measurements provide excellent coverage of the orbit with no large gaps in orbital phase, and the errors in our fit are slightly lower than those of \citet{grundstrom2007} and significantly lower than those of \citet{casares2005a}.  We also emphasize that our final orbital solution relies upon the fixed $P$ from \citet{gregory2002}, but we found very similar values for each orbital element when we allowed $P$ to vary.  We found $P = 26.4982 \pm .0076$ d using the fitting program of \citet{morbey1974} with the combined $V_r$ data, which is not an improvement over \citet{gregory2002}. 

\placefigure{lsi_orbit}


\section{Observations and Radial Velocity Measurements of \ls}

Spectra of \ls~were taken at the CTIO 1.5m telescope between 2007 August and 2008 April in service mode.  We used grating 47 in the second order, a 200 micron slit, the BG39 filter, and the Loral 1K CCD detector.  Spectra were wavelength calibrated using HeAr comparison lamp spectra generally taken before and after each exposure.  The observed wavelength region was therefore about 4050--4700 \AA, with $2500 < R < 3150$ across the chip.

For each night that we had service time available in fall 2007, we obtained 4-6 spectra, each with an exposure time of 30 minutes.  During the spring of 2008, we obtained 2 spectra each night of observations, each of 22 minutes duration.  Thus we obtained a total of 98 blue spectra of \ls.  The images were zero corrected, flat fielded, and wavelength calibrated onto a log wavelength scale using standard procedures in IRAF, and all spectra were interpolated onto a common heliocentric wavelength grid.  The spectra were rectified to a unit continuum using line-free regions.  

Since the signal-to-noise (S/N) ratio was somewhat low, we chose to coadd pairs of consecutive spectra from each night.  The total elapsed time during two consecutive observations is only one hour, about 1\% of the orbital period, so coadding the spectra does not significantly affect the measured $V_r$.  On two nights we collected an odd number of spectra, and in these cases the last spectrum of each set was examined individually.  Therefore, we had 50 new spectra from which to obtain $V_r$ measurements and improve the orbital elements.  

\begin{figure}
\includegraphics[angle=90,scale=0.33]{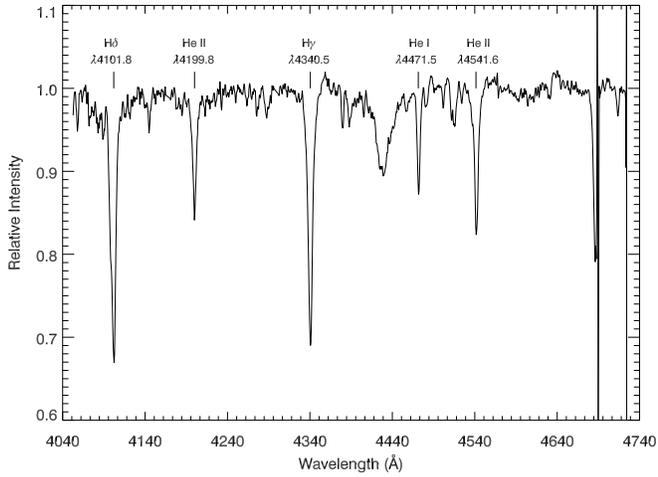}
\\
\caption{The mean spectrum of LS 5039 from the night of 2007 August 29, taken near the time of inferior conjunction of the optical star.  We used this mean spectrum as a reference to determine the relative $V_r$ by cross correlation with each spectrum of LS 5039.  
\label{ls5039_mean} }
\end{figure}

The spectra from August 29, 2007 were taken near the time of inferior conjunction and thus showed the least variation in $V_r$, so we chose to average these six spectra to produce a reference spectrum.  This reference spectrum is plotted in Figure \ref{ls5039_mean}.  Five lines were examined in each spectrum:  H$\gamma$, H$\delta$, \ion{He}{2} $\lambda\lambda$4200, 4542, and \ion{He}{1} $\lambda$4471.  The spectra also recorded the strong \ion{He}{2} $\lambda$4686 line, but a bad column on the chip interfered with measuring reliable $V_r$ from this line.  We performed a cross correlation between each spectrum and the reference spectrum to obtain its relative $V_r$, and then we used the mean $V_r$ from a parabolic fit of the line cores in the reference spectrum to find its absolute $V_r$.  

\placefigure{ls5039_mean}

The period of the system was obtained using the discrete Fourier transform and CLEAN deconvolution program of \citet{roberts1987} (written in IDL\footnote{IDL is a registered trademark of Research Systems, Inc.} by A. W. Fullerton).  The  $V_r$ from our observations were combined with the mean values from all lines measured by \citet{mcswain2001}, \citet{mcswain2004}, and \citet{casares2005b}.  The $V_r$ measurements of \citet{casares2005b} indicate a value of $\gamma$ that is systematically higher than our data sets by 7.0 km~s$^{-1}$, so we subtracted this value from their measurements.  No significant difference was found between the $V_r$ data presented here and past measurements obtained by our group.  We used all of the available $V_r$ with the CLEAN algorithm, which found a strong peak at $P = 3.9060263$ d (which was improved slightly with the later orbital fit), nearly identical to the period found by \citet{casares2005b}.  Our group has previously found $P = 4.1$ d \citep{mcswain2001} and $P = 4.4$ d \citep{mcswain2004} for \ls, due to the very high scatter in its $V_r$ curve, which has lead to considerable confusion about the correct orbital period of this system.  We note that our period search also resulted in $P = 3.906$ d when we excluded the data of \citet{casares2005b}, so we have added confidence that this is indeed the correct period for \ls.  We also investigated other weak frequency signals found by CLEAN, but these periods did not produce a realistic $V_r$ curve.

\begin{figure}
\includegraphics[angle=90,scale=0.33]{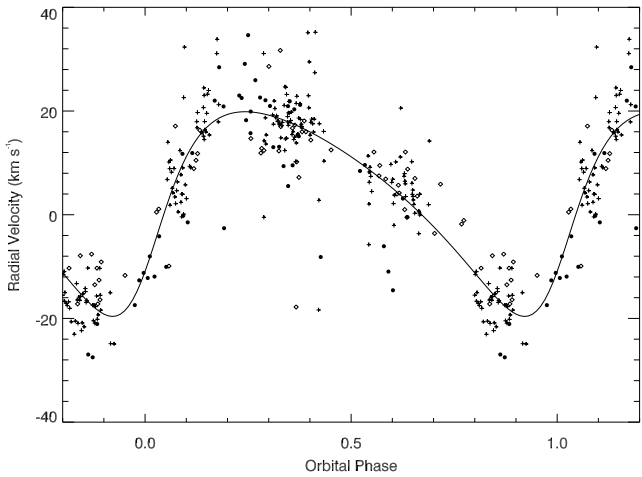}
\\
\caption{Radial velocity curve of LS 5039.  Periastron corresponds to $\phi = 0$ in this plot.  Points plotted as open diamonds are from \citet{mcswain2001, mcswain2004}, crosses are from \citet{casares2005b}, and filled circles are from this work.
\label{vrcurve_ls5039} }
\end{figure}

We determined the remaining orbital elements with the non-linear, least-squares fitting program
of \citet{morbey1974} to solve for the orbital elements.  
With each point assigned equal weight, the resulting orbital elements are displayed in column 3 of Table \ref{orbit}.  Our times of observations, orbital phases, $V_r$ measurements, and $(O-C)$ errors are listed in Table \ref{vr}.  The standard deviation of $V_r$ from the mean of all five lines is comparable to the $(O-C)$ error.  Figure \ref{vrcurve_ls5039} illustrates the radial velocity curve using all available $V_r$ measurements.  

\placefigure{vrcurve_ls5039}

\begin{figure}
\includegraphics[angle=90,scale=0.35]{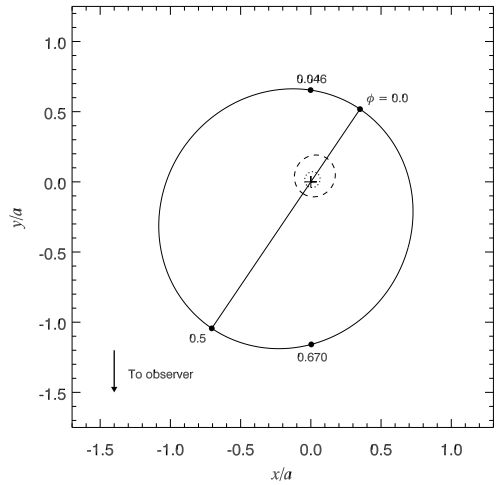} 
\\
\caption{Orbital geometry of \ls, looking down upon the orbital plane, in a similar format to Fig.\ \ref{lsi_orbit}.  Here, the optical star ($23 M_\odot$; \citealt{casares2005b}) orbits either a $3.7 M_\odot$ black hole \citep{casares2005b} or a $1.4 M_\odot$ neutron star.  
\label{ls5039_orbit} }
\end{figure}

For our new orbital fit of \ls, the geometry is illustrated in Figure \ref{ls5039_orbit}.  In this figure, we show the relative orbit of the optical star with $23 M_\odot$ \citep{casares2005b} and a compact companon.  \citet{casares2005b} have proposed a $3.7 M_\odot$ black hole in the system, although \citet{dubus2006} favors a neutron star ($1.4 M_\odot$) companion.  
Our orbital solution is not very different from the fit found by \citet{casares2005b} using all available H, \ion{He}{1}, and \ion{He}{2} lines.  However, they adopt a final solution found using only the \ion{He}{2} lines, which results in a slightly higher $e$, $\gamma$, and $K_1$ than our result.  Their argument that wind emission in the O star may affect the line species differently, whereas the \ion{He}{2} line is formed in the photosphere and is less contaminated by the wind, is very reasonable.  However, we did not observe systematic line-to-line differences in our spectra, and the scatter for all lines is equally large.  Therefore we favor using the mean of all available lines for the orbital solution of \ls.  

\placefigure{ls5039_orbit}

\section{Discussion}

Both \lsi~and \ls~have previously published orbital solutions in the literature with large disagreements in the orbital elements \citep{mcswain2001, mcswain2004, casares2005a, casares2005b, grundstrom2007}.  The need for an improved system geometry, especially in \lsi, was highlighted recently by \citet{sierpowska2008} because the stellar separation and wind density play a significant role in the location of the $\gamma$-ray emission region \citep{dubus2006} and opacities in a pulsar wind zone model.  In fact, they found that a putative wind shock region in \lsi~could lie unphysically close to the optical star using the orbital solution of \citet{casares2005a} along with several assumptions about the stellar and pulsar wind strengths.  Our new solution for \lsi~results in a larger separation distance at periastron, so the stand-off distance of the shock region will be correspondingly farther from the star for the same wind parameters.  Since the $\gamma$-ray production and its opacities are very sensitive to the orbital geometry, our spectroscopic orbital solutions of \lsi~and \ls~may be further improved and the system inclinations constrained using their MeV-TeV light curves.

\begin{figure}
\includegraphics[angle=90,scale=0.33]{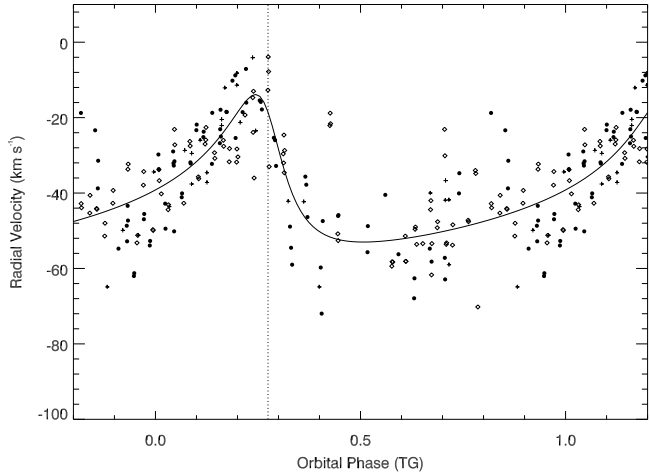} 
\\
\caption{Radial velocity curve of \lsi.  Periastron corresponds to $\phi (\rm TG) = 0.275$ and is marked as a dotted line in this plot.  Points plotted as open diamonds are from \citet{grundstrom2007}, crosses are from \citet{casares2005a}, and filled circles are from this work.
\label{vrcurve_lsi} }
\end{figure}

\begin{figure}
\includegraphics[angle=0,scale=0.5]{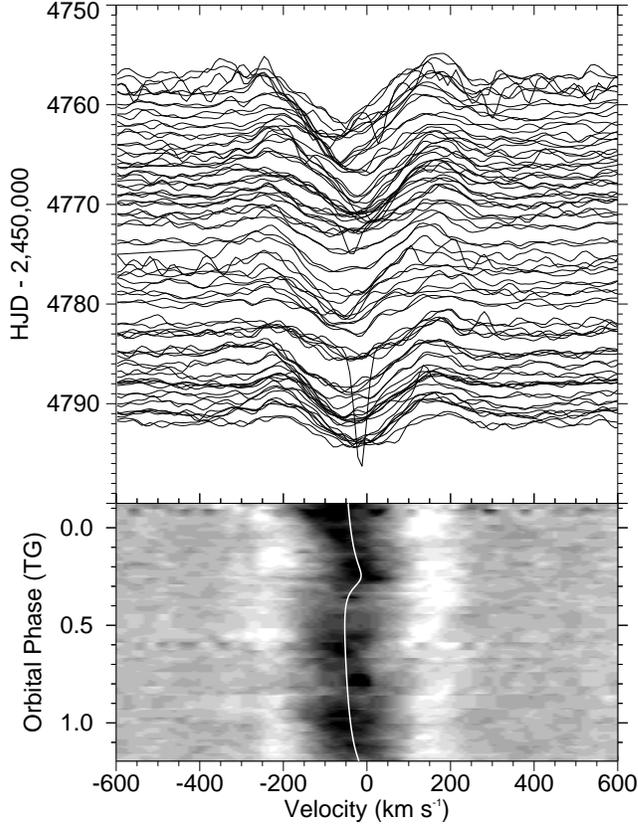}
\caption{The upper plot shows the \ion{He}{1} $\lambda6678$ line profile of \lsi~over our continuous 35 nights of observation, sorted by HJD and smoothed for clarity, and the lower plot shows a gray-scale image of the same line.  Note that the lower plot of the gray-scale spectra are \textit{not} folded by orbital phase but are placed in the same chronological order as the upper plot, with the orbital phases indicated.  The intensity at each velocity in the gray-scale image is assigned one of 16 gray levels based on its value between the minimum (dark) and maximum (bright) observed values.  The intensity between observed spectra is calculated by a linear interpolation between the closest observed phases.  The solid white line shows the theoretical $V_r$ curve solution.  
\label{he6678_gray} }
\end{figure}

We note that the radial velocity curves of both \lsi~and \ls~reveal a large number of spurious $V_r$ points near apastron.  In the case of \lsi, Figure \ref{vrcurve_lsi} indicates a secondary maximum in the $V_r$ curve near $\phi \sim 0.8$ that appears both in our observations and those of \citet{grundstrom2007}.  To examine this behavior more carefully, we show in Figure \ref{he6678_gray} the \ion{He}{1} $\lambda6678$ line profiles and a gray-scale image of this line over our 35 consecutive nights of observation.  Some glitches that do not affect the absorption line profile have been removed for clarity.  Note that neither the line profiles nor gray-scale plots are folded by orbital phase, but rather they reveal true chronological variations in the line profile behavior as a function of HJD and the corresponding orbital phase.  Our spectra between $0.6 \lesssim \phi$(TG)$\lesssim 0.7$ reveal a slight increase in both the red ($R$) and violet ($V$) emission peaks, followed by a decline in emission between $0.7 \lesssim \phi$(TG) $\lesssim 0.9$.  The effect is much more pronounced in the H$\alpha$ line profile and was also noted by \citet{grundstrom2007}.  Thus the $V$ and $R$ variations appear to be related to cyclic orbital variations rather than an isolated event.  Although one spectrum does contain a sharp glitch in the line core (not included in our $V_r$ measurements), the secondary $V_r$ increase near $\phi$(TG) $\sim 0.8$ is due to the decrease in emission strength and the corresponding broadening of the absorption line profile.

\placefigure{he6678_gray}

In \ls, most of the spurious points in the $V_r$ curve occur between $0.4 \lesssim \phi \lesssim 0.6$, and they have been observed by other authors as well \citep{mcswain2001, mcswain2004, casares2005b}.  In fact, the frequent deviations in $V_r$ are the main source of confusion in the orbital elements published by these authors.  Our low spectral resolution does not permit a detailed analysis of the line behavior at this time, but higher resolution spectra may provide evidence of systematic line profile changes in the system.

\acknowledgments
We gratefully acknowledge the anonymous referee for helpful comments about this manuscript as well as Jorge Casares for sharing his radial velocity measurements of LS 5039.  We also thank Di Harmer and the staff at KPNO, as well as Fred Walter and the SMARTS service observers, for their hard work to schedule and support these observations.  This work is supported by NASA DPR numbers NNX08AV70G, NNX08AX79G, and  NNG08E1671 and an institutional grant from Lehigh University.  

{\it Facilities:} \facility{CTIO:1.5m}, \facility{KPNO:CFT}



\begin{deluxetable}{lccccc}
\tablewidth{0pt}
\tablecaption{Radial Velocity Measurements \label{vr} }
\tablehead{
\colhead{ } &
\colhead{HJD} &
\colhead{Orbital} &
\colhead{$V_r$} &
\colhead{$(O-C)$} &
\colhead{$\Delta V_{\rm ISM}$} \\
\colhead{Star} &
\colhead{($-$2,450,000)} &
\colhead{Phase}         &
\colhead{(km s$^{-1}$)} &
\colhead{(km s$^{-1}$)} &
\colhead{(km s$^{-1}$)} }
\startdata
LS~I~$+61^\circ 303$ &  4757.6644   &  0.910 & \phn         $ -54.7$ &         $ -11.1$ 	& \phs $3.92$ 	\\
LS~I~$+61^\circ 303$ &  4758.6607   &  0.947 & \phn         $ -62.0$ &         $ -20.1$ 	& $-2.44$	\\
LS~I~$+61^\circ 303$ &  4758.6827   &  0.948 & \phn         $ -61.3$ &         $ -19.4$ 	& $-0.42$	\\
LS~I~$+61^\circ 303$ &  4759.6877   &  0.986 & \phn         $ -53.9$ &         $ -14.0$ 	& $-0.59$	\\
LS~I~$+61^\circ 303$ &  4759.7088   &  0.987 & \phn         $ -52.7$ &         $ -12.8$ 	& \phs $3.97$	\\
LS~I~$+61^\circ 303$ &  4760.6803   &  0.024 & \phn         $ -42.8$ &\phn     $  -5.1$ 	& $-1.21$	\\
LS~I~$+61^\circ 303$ &  4760.7014   &  0.024 & \phn         $ -49.5$ &         $ -11.8$ 	& $-0.60$	\\
LS~I~$+61^\circ 303$ &  4761.7508   &  0.064 & \phn         $ -41.1$ &\phn     $  -6.3$ 	& \phs $0.21$	\\
LS~I~$+61^\circ 303$ &  4761.7719   &  0.065 & \phn         $ -40.1$ &\phn     $  -5.3$ 	& $-1.38$	\\
LS~I~$+61^\circ 303$ &  4762.6955   &  0.100 & \phn         $ -23.3$ &\phn\phs $   8.4$ 	& \phs $1.87$	\\
LS~I~$+61^\circ 303$ &  4762.7166   &  0.101 & \phn         $ -21.9$ &\phn\phs $   9.8$ 	& \phs $0.74$	\\
LS~I~$+61^\circ 303$ &  4763.7018   &  0.138 & \phn         $ -32.3$ &\phn     $  -4.7$ 	& \phs $2.48$	\\
LS~I~$+61^\circ 303$ &  4763.7229   &  0.139 & \phn         $ -18.7$ &\phn\phs $   8.7$ 	& \phs $1.13$	\\
LS~I~$+61^\circ 303$ &  4764.6373   &  0.173 & \phn         $ -18.4$ &\phn\phs $   4.4$ 	& \phs $2.84$	\\
LS~I~$+61^\circ 303$ &  4764.6583   &  0.174 & \phn         $ -18.5$ &\phn\phs $   4.2$	& \phs $0.74$	 \\
LS~I~$+61^\circ 303$ &  4765.0282   &  0.188 & \phn         $ -10.2$ &\phs     $  10.4$	& \phs $1.82$	 \\
LS~I~$+61^\circ 303$ &  4765.6837   &  0.213 & \phn         $ -18.5$ &\phn     $  -1.8$ 	& \phs $2.87$	\\
LS~I~$+61^\circ 303$ &  4765.9010   &  0.221 & \phn\phn $  -7.1$  &\phn\phs $8.5$ 	& $-0.42$	\\
LS~I~$+61^\circ 303$ &  4765.9221   &  0.222 & \phn         $ -16.0$ &\phn     $  -0.5$ 	& $-2.31$	\\
LS~I~$+61^\circ 303$ &  4766.7893   &  0.254 & \phn         $ -15.4$ &\phn     $  -1.1$ 	& \phs $0.14$	\\
LS~I~$+61^\circ 303$ &  4766.8574   &  0.257 & \phn         $ -15.8$ &\phn     $  -1.2$ 	& $-4.19$	\\
LS~I~$+61^\circ 303$ &  4766.9291   &  0.260 & \phn         $ -17.8$ &\phn     $  -2.8$ 	& $-2.97$	\\
LS~I~$+61^\circ 303$ &  4767.7007   &  0.289 & \phn         $ -25.4$ &\phn     $  -1.6$ 	& \phs $0.01$	\\
LS~I~$+61^\circ 303$ &  4767.7805   &  0.292 & \phn         $ -26.0$ &\phn     $  -1.0$ 	& $-2.20$	\\
LS~I~$+61^\circ 303$ &  4767.8459   &  0.294 & \phn         $ -32.8$ &\phn     $  -6.8$ 	& $-0.05$	\\
LS~I~$+61^\circ 303$ &  4768.7456   &  0.328 & \phn         $ -48.9$ &         $ -10.3$ 	& $-1.69$	\\
LS~I~$+61^\circ 303$ &  4768.8149   &  0.331 & \phn         $ -54.5$ &         $ -15.2$ 	& \phs $0.32$	\\
LS~I~$+61^\circ 303$ &  4768.8841   &  0.333 & \phn         $ -59.0$ &         $ -18.9$ 	& $-1.01$	\\
LS~I~$+61^\circ 303$ &  4769.7368   &  0.365 & \phn         $ -35.6$ &\phs     $  11.0$ 	& $-2.74$	\\
LS~I~$+61^\circ 303$ &  4769.7970   &  0.368 & \phn         $ -37.8$ &\phn\phs $   9.2$ 	& $-0.60$	\\
LS~I~$+61^\circ 303$ &  4769.8738   &  0.371 & \phn         $ -46.4$ &\phn\phs $   1.0$ 	& $-1.05$	\\
LS~I~$+61^\circ 303$ &  4770.7309   &  0.403 & \phn         $ -59.8$ &\phn     $  -9.3$ 	& $-0.65$	\\
LS~I~$+61^\circ 303$ &  4770.7911   &  0.405 & \phn         $ -72.0$ &         $ -21.4$ 	& \phs $2.43$	\\
LS~I~$+61^\circ 303$ &  4770.8624   &  0.408 & \phn         $ -47.4$ &\phn\phs $   3.3$ 	& $-0.92$	\\
LS~I~$+61^\circ 303$ &  4771.8542   &  0.445 & \phn         $ -46.1$ &\phn\phs $   6.2$ 	& \phs $0.64$	\\
LS~I~$+61^\circ 303$ &  4771.8753   &  0.446 & \phn         $ -45.9$ &\phn\phs $   6.5$ 	& $-0.43$	\\
LS~I~$+61^\circ 303$ &  4773.7536   &  0.517 & \phn         $ -55.7$ &\phn     $  -2.7$ 	& $-1.52$	\\
LS~I~$+61^\circ 303$ &  4773.7752   &  0.518 & \phn         $ -48.8$ &\phn\phs $   4.2$ 	& $-0.01$	\\
LS~I~$+61^\circ 303$ &  4774.9024   &  0.560 & \phn         $ -40.5$ &\phs     $  12.2$ 	& $-2.28$	\\
LS~I~$+61^\circ 303$ &  4775.7528   &  0.593 & \phn         $ -56.2$ &\phn     $  -3.9$ 	& $-0.77$	\\
LS~I~$+61^\circ 303$ &  4776.7648   &  0.631 & \phn         $ -67.9$ &         $ -16.2$		& \phs $2.28$	\\
LS~I~$+61^\circ 303$ &  4776.7862   &  0.632 & \phn         $ -62.6$ &         $ -11.0$ 	& $-0.37$	\\
LS~I~$+61^\circ 303$ &  4777.7764   &  0.669 & \phn         $ -54.7$ &\phn     $  -3.8$ 	& \phs $0.17$	\\
LS~I~$+61^\circ 303$ &  4777.8919   &  0.673 & \phn         $ -58.2$ &\phn     $  -7.4$ 	& \phs $0.99$	\\
LS~I~$+61^\circ 303$ &  4778.7466   &  0.706 & \phn         $ -57.2$ &\phn     $  -7.1$ 	& $-1.14$	\\
LS~I~$+61^\circ 303$ &  4778.7677   &  0.706 & \phn         $ -62.9$ &         $ -12.8$ 	& \phs $1.41$	\\
LS~I~$+61^\circ 303$ &  4779.6636   &  0.740 & \phn         $ -40.1$ &\phn\phs $   9.1$ 	& \phs $0.96$	\\
LS~I~$+61^\circ 303$ &  4779.6851   &  0.741 & \phn         $ -34.7$ &\phs     $  14.5$ 	& \phs $1.70$	\\
LS~I~$+61^\circ 303$ &  4781.7344   &  0.818 & \phn         $ -18.8$ &\phs     $  28.2$ 	& \phs $0.87$	\\
LS~I~$+61^\circ 303$ &  4782.6531   &  0.853 & \phn         $ -23.4$ &\phs     $  22.5$ 	& \phs $1.16$	\\
LS~I~$+61^\circ 303$ &  4782.8258   &  0.859 & \phn         $ -38.8$ &\phn\phs $   6.8$ 	& $-1.88$	\\
LS~I~$+61^\circ 303$ &  4782.8470   &  0.860 & \phn         $ -31.4$ &\phs     $  14.1$ 	& $-0.54$	\\
LS~I~$+61^\circ 303$ &  4784.6952   &  0.930 & \phn         $ -48.7$ &\phn     $  -5.9$ 	& $-2.14$	\\
LS~I~$+61^\circ 303$ &  4784.7163   &  0.931 & \phn         $ -52.8$ &         $ -10.1$ 	& $-0.96$	\\
LS~I~$+61^\circ 303$ &  4784.7536   &  0.932 & \phn         $ -43.4$ &\phn     $  -0.7$ 	& \phs $1.90$	\\
LS~I~$+61^\circ 303$ &  4784.7747   &  0.933 & \phn         $ -47.3$ &\phn     $  -4.7$ 	& $-2.93$	\\
LS~I~$+61^\circ 303$ &  4785.8022   &  0.972 & \phn         $ -47.0$ &\phn     $  -6.2$ 	& $-1.47$	\\
LS~I~$+61^\circ 303$ &  4785.8233   &  0.973 & \phn         $ -45.5$ &\phn     $  -4.9$ 	& $-3.56$	\\
LS~I~$+61^\circ 303$ &  4786.7401   &  0.007 & \phn         $ -29.8$ &\phn\phs $   9.0$ 	& \phs $2.78$	\\
LS~I~$+61^\circ 303$ &  4786.7612   &  0.008 & \phn         $ -33.7$ &\phn\phs $   5.0$ 	& $-0.71$	\\
LS~I~$+61^\circ 303$ &  4787.6918   &  0.043 & \phn         $ -29.0$ &\phn\phs $   7.4$ 	& \phs $0.50$	\\
LS~I~$+61^\circ 303$ &  4787.7129   &  0.044 & \phn         $ -32.1$ &\phn\phs $   4.2$ 	& \phs $2.14$	\\
LS~I~$+61^\circ 303$ &  4787.7340   &  0.045 & \phn         $ -32.7$ &\phn\phs $   3.6$ 	& $-0.47$	\\
LS~I~$+61^\circ 303$ &  4787.7551   &  0.046 & \phn         $ -50.2$ &         $ -13.9$ 	& \phs $1.13$	\\
LS~I~$+61^\circ 303$ &  4787.7762   &  0.046 & \phn         $ -31.6$ &\phn\phs $   4.6$ 	& $-1.42$	\\
LS~I~$+61^\circ 303$ &  4788.8069   &  0.085 & \phn         $ -31.9$ &\phn\phs $   1.2$ 	& $-3.15$	\\
LS~I~$+61^\circ 303$ &  4788.8280   &  0.086 & \phn         $ -32.2$ &\phn\phs $   0.8$ 	& $-3.10$	\\
LS~I~$+61^\circ 303$ &  4789.6443   &  0.117 & \phn         $ -25.2$ &\phn\phs $   4.8$ 	& \phs $2.02$	\\
LS~I~$+61^\circ 303$ &  4789.6654   &  0.118 & \phn         $ -23.7$ &\phn\phs $   6.1$ 	& \phs $1.22$	\\
LS~I~$+61^\circ 303$ &  4790.7113   &  0.157 & \phn         $ -26.8$ &\phn     $  -1.7$ 	& $-1.50$	\\
LS~I~$+61^\circ 303$ &  4790.7324   &  0.158 & \phn         $ -23.1$ &\phn\phs $   1.9$ 	& \phs $0.90$	\\
LS~I~$+61^\circ 303$ &  4790.7536   &  0.159 & \phn         $ -17.9$ &\phn\phs $   6.9$ 	& $-1.35$	\\
LS~I~$+61^\circ 303$ &  4790.7747   &  0.159 & \phn         $ -25.0$ &\phn     $  -0.3$ 	& $-1.89$	\\
LS~I~$+61^\circ 303$ &  4791.7059   &  0.195 & \phn\phn     $  -8.8$ &\phs     $  10.7$ 	& $-1.40$	\\
LS~I~$+61^\circ 303$ &  4791.7270   &  0.195 & \phn         $ -25.4$ &\phn     $  -6.0$ 	& $-1.92$ \\
LS 5039  \dotfill  &   4315.5746 &  0.351 & \phn\phs     $  21.9$ &\phn\phs $   4.1$   &  \nodata \\
LS 5039  \dotfill  &   4315.6176 &  0.362 & \phn\phs     $  20.3$ &\phn\phs $   2.9$   &  \nodata \\
LS 5039  \dotfill  &   4315.6607 &  0.373 & \phn\phs     $  15.1$ &\phn     $  -1.9$   &  \nodata \\
LS 5039  \dotfill  &   4316.6470 &  0.626 & \phn\phn\phs $   3.1$ &\phn     $  -0.3$   &  \nodata \\
LS 5039  \dotfill  &   4316.6900 &  0.637 & \phn\phn     $  -0.5$ &\phn     $  -3.1$   &  \nodata \\
LS 5039  \dotfill  &   4330.4876 &  0.169 & \phn\phs     $  22.0$ &\phn\phs $   3.9$   &  \nodata \\
LS 5039  \dotfill  &   4330.5296 &  0.180 & \phn\phs     $  28.4$ &\phn\phs $   9.8$   &  \nodata \\
LS 5039  \dotfill  &   4330.5716 &  0.190 & \phn\phs     $  20.9$ &\phn\phs $   1.8$   &  \nodata \\
LS 5039  \dotfill  &   4333.6362 &  0.975 & \phn         $ -17.4$ &\phn     $  -1.8$   &  \nodata \\
LS 5039  \dotfill  &   4333.6782 &  0.986 & \phn         $ -12.7$ &\phn\phs $   1.2$   &  \nodata \\
LS 5039  \dotfill  &   4333.7202 &  0.997 & \phn         $ -11.2$ &\phn\phs $   0.5$   &  \nodata \\
LS 5039  \dotfill  &   4334.6263 &  0.229 & \phn\phs     $  23.0$ &\phn\phs $   3.2$   &  \nodata \\
LS 5039  \dotfill  &   4334.6788 &  0.242 & \phn\phs     $  29.1$ &\phn\phs $   9.2$   &  \nodata \\
LS 5039  \dotfill  &   4334.7103 &  0.250 & \phn\phs     $  34.6$ &\phs     $  14.8$   &  \nodata \\
LS 5039  \dotfill  &   4338.5552 &  0.234 & \phn\phs     $  22.5$ &\phn\phs $   2.7$   &  \nodata \\
LS 5039  \dotfill  &   4338.5973 &  0.245 & \phn\phs     $  18.2$ &\phn     $  -1.6$   &  \nodata \\
LS 5039  \dotfill  &   4338.6393 &  0.256 & \phn\phs     $  15.7$ &\phn     $  -4.1$   &  \nodata \\
LS 5039  \dotfill  &   4341.5908 &  0.011 & \phn\phn     $  -8.0$ &\phn\phs $   0.2$   &  \nodata \\
LS 5039  \dotfill  &   4341.6354 &  0.023 & \phn         $ -11.9$ &\phn     $  -6.4$   &  \nodata \\
LS 5039  \dotfill  &   4341.6797 &  0.034 & \phn\phn     $  -4.2$ &\phn     $  -1.5$   &  \nodata \\
LS 5039  \dotfill  &   4342.5470 &  0.256 & \phn\phs     $  19.9$ &\phn\phs $   0.1$   &  \nodata \\
LS 5039  \dotfill  &   4342.5912 &  0.268 & \phn\phs     $  25.9$ &\phn\phs $   6.2$   &  \nodata \\
LS 5039  \dotfill  &   4342.6361 &  0.279 & \phn\phs     $  22.6$ &\phn\phs $   3.0$   &  \nodata \\
LS 5039  \dotfill  &   4343.5828 &  0.521 & \phn\phn\phs $   8.4$ &\phn     $  -1.5$   &  \nodata \\
LS 5039  \dotfill  &   4343.6293 &  0.533 & \phn\phn\phs $   9.6$ &\phn\phs $   0.3$   &  \nodata \\
LS 5039  \dotfill  &   4343.6734 &  0.545 & \phn\phn\phs $   8.2$ &\phn     $  -0.4$   &  \nodata \\
LS 5039  \dotfill  &   4346.5929 &  0.292 & \phn\phs     $  22.1$ &\phn\phs $   2.7$   &  \nodata \\
LS 5039  \dotfill  &   4346.6358 &  0.303 & \phn\phs     $  21.0$ &\phn\phs $   1.8$   &  \nodata \\
LS 5039  \dotfill  &   4346.6673 &  0.311 & \phn\phs     $  13.0$ &\phn     $  -5.9$   &  \nodata \\
LS 5039  \dotfill  &   4354.5347 &  0.325 & \phn\phs     $  13.1$ &\phn     $  -5.5$   &  \nodata \\
LS 5039  \dotfill  &   4354.5767 &  0.336 & \phn\phn\phs $   9.4$ &\phn     $  -8.9$   &  \nodata \\
LS 5039  \dotfill  &   4354.6187 &  0.347 & \phn\phs     $  20.9$ &\phn\phs $   3.0$   &  \nodata \\
LS 5039  \dotfill  &   4355.5278 &  0.580 & \phn\phn     $  -6.1$ &         $ -12.6$   &  \nodata \\
LS 5039  \dotfill  &   4355.5705 &  0.590 & \phn         $ -11.0$ &         $ -16.7$   &  \nodata \\
LS 5039  \dotfill  &   4355.6133 &  0.601 & \phn         $ -14.6$ &         $ -19.6$   &  \nodata \\
LS 5039  \dotfill  &   4357.5744 &  0.103 & \phn\phn     $  -1.5$ &         $ -13.2$   &  \nodata \\
LS 5039  \dotfill  &   4357.6174 &  0.114 & \phn\phs     $  11.9$ &\phn     $  -1.4$   &  \nodata \\
LS 5039  \dotfill  &   4357.5314 &  0.092 & \phn\phn     $  -0.2$ &         $ -10.2$   &  \nodata \\
LS 5039  \dotfill  &   4358.6397 &  0.376 & \phn\phs     $  21.2$ &\phn\phs $   4.3$   &  \nodata \\
LS 5039  \dotfill  &   4358.5687 &  0.358 & \phn\phn\phs $   9.5$ &\phn     $  -8.0$   &  \nodata \\
LS 5039  \dotfill  &   4358.5263 &  0.347 & \phn\phn\phs $   5.5$ &         $ -12.4$   &  \nodata \\
LS 5039  \dotfill  &   4360.5797 &  0.873 & \phn         $ -27.5$ &\phn     $  -9.9$   &  \nodata \\
LS 5039  \dotfill  &   4360.5373 &  0.862 & \phn         $ -27.0$ &         $ -10.2$   &  \nodata \\
LS 5039  \dotfill  &   4360.6222 &  0.884 & \phn         $ -21.1$ &\phn     $  -2.7$   &  \nodata \\
LS 5039  \dotfill  &   4525.8807 &  0.192 & \phn\phn     $  -2.6$ &         $ -21.7$   &  \nodata \\
LS 5039  \dotfill  &   4536.8753 &  0.006 & \phn         $ -12.2$ &\phn     $  -2.7$   &  \nodata \\
LS 5039  \dotfill  &   4544.8626 &  0.051 & \phn         $ -10.1$ &         $ -11.6$   &  \nodata \\
LS 5039  \dotfill  &   4552.8287 &  0.091 & \phn\phs     $  11.8$ &\phn\phs $   2.1$   &  \nodata \\
LS 5039  \dotfill  &   4560.8249 &  0.138 & \phn\phs     $  15.8$ &\phn     $  -0.0$   &  \nodata \\
LS 5039  \dotfill  &   4569.7635 &  0.426 & \phn\phn     $  -8.2$ &         $ -23.0$   &  \nodata \\
\enddata
\end{deluxetable}

\begin{deluxetable}{lcc}
\tablewidth{0pt}
\tablecaption{Orbital Elements \label{orbit}}
\tablehead{
\colhead{ } &
\colhead{\lsi} &
\colhead{\ls} }
\startdata
$P$ (d) \dotfill				&  26.4960\tablenotemark{a}	&  $3.90608 \pm  .00010$	\\
$T_0$ (HJD--2,450,000) \dotfill	&  $1057.89 \pm  .23$		&  $2825.985 \pm .053$	\\
$e$ 	\dotfill				&  $.537 \pm .034$			&  $.337 \pm .036$	\\
$\omega$ (deg) \dotfill		&  $40.5 \pm 5.7$			&  $236.0 \pm  5.8$	\\
$K_1$ (km s$^{-1}$)	 \dotfill	&  $19.6 \pm 1.1$			&  $19.74 \pm .86$	\\
$\gamma$ (km s$^{-1}$) \dotfill	&  $-41.41 \pm .60$			&  $4.01 \pm  .31$	\\
$f(m)$ ($M_\odot$) \dotfill		&  $0.0124 \pm .0022$		&  $.00261 \pm  .00036$	\\
$a_1 \sin i$ ($R_\odot$) \dotfill	&  $8.64 \pm 0.52$			&  $1.435 \pm  0.066$	\\
$\sigma$ (km s$^{-1}$) \dotfill	&  $7.41$					&  $7.14$	\\
\enddata
\tablenotetext{a}{Fixed.}
\end{deluxetable}

\clearpage

\clearpage

\clearpage

\end{document}